\documentclass[12pt]{article}

\usepackage[numbers,sort]{natbib}
\usepackage{amsmath}
\usepackage{amssymb}
\usepackage{hyperref}

\usepackage{sectsty}
\sectionfont{\large}
\subsectionfont{\normalsize}

\newcommand{\ba}{\begin{aligned}}
\newcommand{\ea}{\end{aligned}}
\def\ben{\begin{equation*}}
\def\een{\end{equation*}}
\newcommand{\beqs}{\begin{eqnarray}}
\newcommand{\eeqs}{\end{eqnarray}}

\newcommand{\sdot}{\hspace{-3pt}\cdot\hspace{-3pt}}

\def\be{\begin{equation}}
\def\ee{\end{equation}}
\def\bea{\begin{eqnarray}}
\def\eea{\end{eqnarray}}
\def\bsp{\be\begin{split}}

\def\d{\delta}
\def\e{\epsilon}
\def\m{\mu}

\def\n{\nu}
\def\s{\sigma}

\def\l{\lambda}

\makeatletter

\newcommand{\Rmnum}[1]{\expandafter\@slowromancap\romannumeral #1@}
\makeatother

\renewcommand{\title}[1]{\vbox{\center\LARGE{#1}}\vspace{5mm}}
\renewcommand{\author}[1]{\vbox{\center\large{#1}}\vspace{5mm}}

\newcommand{\pd}{\partial}

\everymath{\displaystyle}

\begin{document}

\begin{titlepage}
\vspace{10pt} \hfill { ICTP-SAIFR/2014-005} \vspace{20mm}
\begin{center}

{\Large \bf Next to subleading soft-graviton theorem in arbitrary dimensions}


\vspace{45pt}

{
\textbf{Chrysostomos Kalousios},$^a$ \textbf{Francisco Rojas},$^b$
\footnote[1]{\href{mailto:ckalousi@ift.unesp.br}{\tt{ckalousi@ift.unesp.br}}, \href{mailto:frojasf@ift.unesp.br}{\tt{frojasf@ift.unesp.br}}}
}
\\[15mm]

{\it\ ${}^a\,$ICTP South American Institute for Fundamental Research\\
Instituto de F\'\i sica Te\'orica, UNESP-Universidade Estadual Paulista\\
R. Dr. Bento T. Ferraz 271 - Bl. II, 01140-070, S\~ao Paulo, SP, Brasil}\\
\vspace{10pt}
{\it\ ${}^b\,$Instituto de F\'\i sica Te\'orica, UNESP-Universidade Estadual Paulista\\
R. Dr. Bento T. Ferraz 271 - Bl. II, 01140-070, S\~ao Paulo, SP, Brasil}

\vspace{20pt}

\end{center}

\vspace{40pt}

\centerline{{\bf{Abstract}}}
\vspace*{5mm}
\noindent
We study the soft graviton theorem recently proposed by Cachazo and Strominger.  We employ the Cachazo, He and Yuan formalism to show that the next to subleading order soft factor for gravity is universal at tree level in arbitrary dimensions.
\vspace{15pt}
\end{titlepage}

\newpage

\section{Introduction}

The study of the soft graviton amplitudes dates back to Weinberg \cite{Weinberg:1964ew, Weinberg:1965nx} where the leading soft behavior was obtained. In \cite{Strominger:2013jfa,He:2014laa,Kapec:2014opa} a new soft graviton theorem, conjectured to be the Ward identities of a new symmetry of the quantum gravity S-matrix,\footnote{This new proposed symmetry is an extension of the Bondi, van der Burg, Metzner and Sachs (BMS) symmetry \cite{Bondi:1962px,Sachs:1962wk}.} was proposed. Cachazo and Strominger \cite{Cachazo:2014fwa} have recently shown that the new conjectured soft behavior, through subleading and next-to-subleading orders in the soft expansion, has a universal form in four spacetime dimensions at tree level.\footnote{Early results for soft photons at subleading order were obtained in \cite{Low:1958sn,Burnett:1967km,Bell:1969yw,DelDuca:1990gz}. Gross and Jackiw, using dispersion relation methods, derived the subleading soft factor for graviton scattering off scalars in \cite{Gross:1968in}, and White revised the subject in \cite{White:2011yy} using path integral resummation techniques.} An extension to gluons for the first subleading soft behavior at tree level was reported in \cite{Casali:2014xpa}. Using Feynman diagram techniques, the first subleading theorem was also confirmed in \cite{White:2014qia}. In \cite{Larkoski:2014hta} it was demonstrated that the conformal invariance of tree level gauge theory amplitudes in four spacetime dimensions determines the form of the first subleading theorem. 

Very recently it has been shown that the form of the first subleading term in the soft expansion in $D$ dimensions is highly constrained by the requirements from Poincar\'e symmetry and gauge invariance \cite{Broedel:2014fsa}. Subsequently, the authors of \cite{Bern:2014vva} have shown that on-shell gauge invariance determines the complete form of the first two subleading soft graviton theorems. Using the Cachazo, He, Yuan (CHY) formula \cite{Cachazo:2013hca}, the universality of the soft behavior to first subleading order has been shown to hold in $D$ dimensions \cite{Schwab:2014xua, Afkhami-Jeddi:2014fia}.  The purpose of the present note is to use the CHY formula to prove the universal nature of the next-to-subleading soft graviton theorem at tree level in arbitrary dimensions.

Studies on loop corrections to subleading soft theorems have been presented in \cite{Bern:2014oka, He:2014bga, Cachazo:2014dia}. Progress in the context of string theory has been reported in \cite{Schwab:2014fia,Bianchi:2014gla} and also in \cite{Adamo:2014yya,Geyer:2014lca} relevant for recent twistor constructions. More recent progress on soft theorems in the context of massless QED has appeared in \cite{He:2014cra,Lysov:2014csa}.

The conjecture of \cite{Cachazo:2014fwa} states, for an on-shell tree level $n$-graviton amplitude $M_n$, that
\be\label{CS_conjecture}
M_n = \left( \frac 1 \lambda S^{(0)} + S^{(1)} + \lambda S^{(2)} + \mathcal{O}(\lambda^2) \right) M_{n-1},
\ee
where $n$ is taken to be the soft particle with momentum $k_n$ and we scale the momentum $k_n \to \lambda k_n$ and take the limit when $\lambda$ approaches zero.  In the above, 
\be 
\label{S0}
S^{(0)} = \sum_{a=1}^{n-1} \frac{\epsilon_{\mu\nu} k_a^{\mu} k_a^{\nu}}{k_n \sdot k_a}
\ee
is Weinberg's soft theorem with $\epsilon_{\mu\nu}$ denoting the polarization tensor of the soft graviton and the gravitational constant has been set to 1.  The conjectured forms of the subleading and next-to-subleading theorems are
\be\label{S2}
S^{(1)} = -i \sum_{a=1}^{n-1} \frac{\epsilon_{\m\n} k_a^{\mu} k_{n\l} J_a^{\l\n}} {k_n \sdot k_a}, \qquad
S^{(2)} = - \frac 1 2 \sum_{a=1}^{n-1} \frac{\epsilon_{\m\n} k_{n\rho}J_a^{\rho\m}k_{n\l}J_a^{\l\n}} {k_n \sdot k_a}.
\ee
In order to treat gluon and graviton polarizations on an equal footing one can choose to write the graviton polarization for the $a^{\rm th}$ particle as
\be
\e_{a\m\n} = \e_{a\m} \e_{a\n}
\ee
where $a=1, \dots, n-1$. Tracelessness and orthogonality to $k_a$ translate into $\e_a \!\cdot \!\e_a =0$ and $\e_a \sdot k_a =0$ respectively.\footnote{We do not use any other gauge condition in this work.}

The subleading contributions to the soft theorem depend on the total angular momentum operator, which is \footnote{We follow the convention $A_{(\mu}B_{\nu)}= A_{\mu}B_{\nu} + A_{\nu}B_{\mu}$ and $A_{[\mu}B_{\nu]}= A_{\mu}B_{\nu} - A_{\nu}B_{\mu}$.}
\be
J_a^{\m\n} =i\left( k_a^{[\m} \frac{\partial}{\partial k_{a\n]}}
			+\e_a^{[\m} \frac{\partial}{\partial \e_{a\n]}} \right)
\ee
for the $a^{\mathrm{th}}$ particle.  Note that in using this formula one should consider the polarization vectors $\e_a^\m$ to be independent of the momenta $k_a^\m$. 

This paper is organized as follows. In Section 2 we review the CHY formalism \cite{Cachazo:2013hca} 
for tree level graviton amplitudes which is valid in arbitrary dimensions and, in this language, we 
set up the computation for the expansion of the amplitude up to next-to-subleading order in the soft parameter. We finish this section by stating the new soft theorem extended to $D$ dimensions. In Section 3 we explicitly evaluate the tree level $n$-graviton amplitude at next-to-subleading order in the soft expansion. In Section 4 we compute the action of the conjectured $S^{(2)}$ operator \eqref{S2} onto the $(n-1)$-graviton amplitude, as stated in \eqref{CS_conjecture}, and show that it perfectly matches with the next-to-subleading amplitude $M^{(2)}$ of Section 3, thus proving the theorem.

\section{Review and setup of the problem}
In this section we briefly review the CHY construction \cite{Cachazo:2013hca} for tree level graviton amplitudes. A key object is the \emph{scattering equations}
\be\label{scat_eqs}
\sum_{b\neq a}^n \frac{k_a \sdot k_b}{\s_{ab}}=0, \quad {a,b=1,\dots, n.}
\ee
with $\s_{ab} \equiv \s_a - \s_b$, where the $\s_a$ are in general complex valued quantities. Due to the $SL(2,\mathbb{C})$ symmetry of \eqref{scat_eqs}, these constitute a system of $n-3$ independent equations for the set $\{\s_a\}$ and one can arbitrarily fix three of the $\s_a$ variables. We will call $\s_i,\s_j, \s_k$ the three fixed $\s$s. The gauge fixed amplitude is
\be\label{gauge_fixed_amplitude}
M_n = \int [d\s]_{n-4} \, d\s_n \prod_{a\neq i,j,k}^n \delta(f_a^n) \, E_n,
\ee
where we have employed the useful short notation
\be
\quad f_a^n \equiv \sum_{b\neq a}^n \frac{k_a \sdot k_b}{\sigma_{ab}},\quad
[d\s]_{n-4} \equiv (\s_{pq} \s_{qr} \s_{rp})(\s_{ij}\s_{jk}\s_{ki}) \prod_{c\neq p,q,r}^{n-1} d\s_c.
\ee
In the above, $E_n$ is defined to be
\be\label{E_n}
E_n = 4 \, \mathrm{det} (\Psi_{xy}^{xy})/\s_{xy}^2 ,
\ee
where $\Psi^{xyz\ldots}_{xyz'\ldots}$ is obtained from the $2n \times 2n$ antisymmetric matrix $\Psi$ after removing rows $x,y,z,\ldots$ and columns $x,y,z',\ldots$ with $1\leq x < y \leq n$.  The explicit expression of $\Psi$ is given by 
\be
\Psi = 
\begin{pmatrix}
A & -C^{\mathrm{T}} \\
C & B
\end{pmatrix}
\ee
with the $n \times n$ matrices $A,B,C$ given by 
\be\ba
A_{ab}=\frac{k_a \sdot k_b}{\s_{ab}} \d_{a\neq b}, \quad B_{ab}=\frac{\e_a \sdot \e_b}{\s_{ab}} \d_{a\neq b}, \quad C_{ab}=\frac{\e_a \sdot k_b}{\s_{ab}} \d_{a\neq b} - \d_{ab} \sum_{c\neq a}^n \frac{\e_a \sdot k_c}{\s_{ac}},
\ea\ee
where we use $\d_{a\neq b}\equiv 1-\d_{ab}$ in order to avoid cluttering our equations. In \cite{Cachazo:2013hca} it was shown that the quantity $E_n$ is independent of the choice of $x$ and $y$.

In order to expand the delta function appearing in \eqref{gauge_fixed_amplitude} in powers of $\l$ we separate it into two parts
\be\ba\label{expansion_of_delta}
\prod_{a \neq i,j,k}^n \delta(f_a^n) &= \frac 1 \lambda \d\left( \sum_{b=1}^{n-1} \frac{k_n \sdot k_b}
     {\s_{nb}} \right) 
\prod_{a\neq i,j,k}^{n-1} \delta\left(\sum_{b\neq a}^{n-1} \frac{k_a \sdot k_b}{\sigma_{ab}}+
     \lambda \frac{k_a \sdot k_n}{\sigma_{an}}\right)\\
&= \delta(f_n^{n-1}) 
	\left(\frac 1 \l \delta^{(0)}+ \delta^{(1)}+
	\lambda \delta^{(2)}\right) +\mathcal{O}(\lambda^2),
\ea\ee
where we define
\be 
\delta^{(0)}=\prod_{a\neq i,j,k}^{n-1}\delta(f_a^{n-1}), \quad 
\delta^{(1)}=\sum_{l\neq i,j,k}^{n-1}\frac{k_l \sdot k_n}{\sigma_{ln}}\delta'(f_l^{n-1})
\left[\prod_{a\neq i,j,k,l}^{n-1}\delta(f_a^{n-1})\right],
\ee
\be\ba\label{delta_2}
\delta^{(2)}=\frac 1 2 \sum_{l\neq i,j,k}^{n-1}
  \frac{k_l \sdot k_n}{\sigma_{ln}}&\delta'(f_l^{n-1})\hspace{-10pt}
  \sum_{m\neq i,j,k,l}^{n-1}\left[\frac{k_m \sdot k_n}
  {\sigma_{mn}}\delta'(f_m^{n-1})\hspace{-10pt}
  \prod_{b\neq i,j,k,l,m}^{n-1}\hspace{-10pt}\delta(f_b^{n-1})\right]\\
&+\frac 1 2 \sum_{l\neq i,j,k}^{n-1}\left(\frac{k_l\sdot k_n}
  {\sigma_{ln}}\right)^2    \delta''(f_l^{n-1})\hspace{-10pt}
  \prod_{b\neq i,j,k,l}^{n-1}\hspace{-10pt}\delta(f_b^{n-1}).
\ea\ee
We also need to expand $E_n$ in \eqref{gauge_fixed_amplitude} to second order in $\l$
\be\label{expansion_of_E}
E_n = E_n^{(0)} + \l E_n^{(1)} + \l^2 E_n^{(2)} + \mathcal{O}(\l^2).
\ee
Plugging \eqref{expansion_of_delta} and \eqref{expansion_of_E} into \eqref{gauge_fixed_amplitude} we get
\be
M_n =\frac 1 \l  M_n^{(0)} + M_n^{(1)} + \l M_n^{(2)} + \mathcal{O}(\l^2),
\ee
where
\be\ba\label{M_n_2}
M_n^{(0)} &= \int [d\s]_{n-4} \, d\s_n \delta(f_n^{n-1}) \d^{(0)} E_n^{(0)}, \quad
M_n^{(1)} = \int [d\s]_{n-4} \, d\s_n \delta(f_n^{n-1}) (\d^{(1)} E_n^{(0)} + \d^{(0)} E_n^{(1)}),\\
M_n^{(2)} &= \int [d\s]_{n-4} \, d\s_n \delta(f_n^{n-1})(\d^{(2)} E_n^{(0)} + \d^{(1)} E_n^{(1)}+\d^{(0)} E_n^{(2)}).
\ea\ee
The soft theorem conjectures that the following equality should hold
\be\label{expected_result}
M_n^{(i)} = S^{(i)} M_{n-1}, \qquad i=0,1,2.
\ee
Weinberg's soft theorem, i.e., $M_n^{(0)}=S^{(0)}M_{n-1}$, can be derived as follows. To evaluate $M_n^{(0)}$ in \eqref{M_n_2} we also need $E_n^{(0)}$, the leading contribution to the determinant \eqref{E_n}, which is $E_n^{(0)} = C_{nn}^2 E_{n-1}$. In order to see that, we can set $\l=0$  in $E_n$.  Then all the elements of the $(n-2)^{\mathrm{th}}$ row vanish apart from the last one which equals $-C_{nn}$.  Similarly all elements of the $(n-2)^{\mathrm{th}}$ column are zero apart from the last one which is $C_{nn}$.  Expansion of the determinant along the aforementioned row and column will yield another extra sign which completes the proof.   

Separating all the dependence on $\s_n$ in $M_n^{(0)}$, i.e.,
\be\ba\label{M_n_0_eval}
M_n^{(0)} &=  \int [d\s]_{n-4} \, \d^{(0)} E_{n-1} \int d\s_n \delta(f_n^{n-1}) C_{nn}^2,
\ea\ee
we can explicitly evaluate the integral over $\s_n$. Due to the absence of branch-cuts and the regularity of the integrand when $\s_n\to\infty$, we may treat the delta function as a pole and we can evaluate the integral by deforming the contour and using the residue theorem. Performing this one obtains
\be\ba
\int d\s_n \delta(f_n^{n-1}) C_{nn}^2=\sum_{a=1}^{n-1} \frac{(\e_n \sdot k_a)^2}{k_n \sdot k_a}.
\ea\ee
Putting everything together into \eqref{M_n_0_eval} yields
\be\ba
\label{M_n_0_eval2}
M_n^{(0)} &=  \sum_{a=1}^{n-1} \frac{(\e_n \sdot k_a)^2}{k_n \sdot k_a} \int [d\s]_{n-4}  \hspace{-5pt}\prod_{l\neq i,j,k}^{n-1}\delta(f_l^{n-1}) E_{n-1} \\
&= \sum_{a=1}^{n-1} \frac{(\e_n \sdot k_a)^2}{k_n \sdot k_a} M_{n-1}.
\ea\ee
From \eqref{S0} one can easily see that $S^{(0)}M_{n-1}$ is precisely the last line of \eqref{M_n_0_eval2}, thus proving Weinberg's leading soft-graviton theorem.

The computation of \eqref{expected_result} for $i=1$ in arbitrary dimensions was performed in \cite{Schwab:2014xua,Afkhami-Jeddi:2014fia}.  In the next section we start the computation of the next to subleading soft contribution ($i=2$) by evaluating $M_n^{(2)}$ in \eqref{M_n_2}.  Then, in Section 4, we will evaluate the action of $S^{(2)}$ on $M_{n-1}$. We will compare both sides of \eqref{expected_result} by matching terms that contain the same support from the $\d$-distributions and we will find perfect matching, thus, proving the theorem. 

\section{Evaluation of $M_n^{(2)}$}
We split the evaluation of $M_n^{(2)}$ into three parts
\be
M_n^{(2)} = \int \, [d\s]_{n-4}\left(m_1 + m_2 + m_3\right), \qquad m_i = \int d\s_n \,\delta(f_n^{n-1})\d^{(3-i)} E_n^{(i-1)}.
\ee
\subsection{Evaluation of $m_1$}
Using \eqref{delta_2}, the first contribution, $m_1$, to $M_n^{(2)}$ is
\be\ba
\label{m1} 
m_1 &= \frac 1 2 E_{n-1} \sum_{l \neq i,j,k}^{n-1} \hspace{0pt}\d'(f_l^{n-1})
                \sum_{m\neq i,j,k,l}^{n-1} \hspace{0pt}\d'(f_m^{n-1}) 
                \prod_{b\neq i,j,k,l,m}^{n-1} \hspace{0pt}\d(f_b^{n-1}) \hspace{5pt} I_1 \\
      &\qquad +\frac 1 2  E_{n-1} \sum_{l \neq i,j,k}^{n-1} \hspace{0pt}\d''(f_l^{n-1})
                \prod_{b\neq i,j,k,l}^{n-1} \hspace{0pt}\d(f_b^{n-1})\hspace{5pt} I_2,
\ea\ee
where we have isolated the integration over $\s_n$ to the following integral
\be\ba\label{I}
I &= k_l \sdot k_n\,k_m \sdot k_n \int d\s_n  \d(f_n^{n-1}) \frac{C_{nn}^2}{\s_{nl}\s_{nm}}. 
\ea\ee
Therefore, in \eqref{m1}, we have $I_1= I|_{m\neq l}$ and $I_2=I|_{m=l}$.

We now move on to compute the integral \eqref{I}. We find
\be\ba\label{Ieval}
I &= \Bigg\{
\left[\frac{k_m \sdot k_n\,\e_n \sdot k_l}{\s_{ml}} 
\left( \frac{\e_n \sdot k_l}{k_n\sdot k_l} \sum_{c\neq l}^{n-1} \frac{k_n \sdot k_c}{\s_{lc}}
        -2 \sum_{c\neq l}^{n-1} \frac{\e_n \sdot k_c}{\s_{lc}}
\right) -\frac{(\e_n \sdot k_l)^2 k_m\sdot k_n}{\s_{ml}^2}\right]
+ (l\leftrightarrow m)\\
& \hspace{1cm} + k_l \sdot k_n\,k_m \sdot k_n \sum_{c\neq l,m}^{n-1} \frac{(\e_n \sdot k_c)^2}{\s_{lc}\s_{mc}k_n\sdot k_c} \Bigg\}\d_{m\neq l}\\
&\quad +\Bigg\{(k_l \sdot k_n)^2 \sum_{c\neq l}^{n-1} \frac{(\e_n \sdot k_c)^2}{\s_{lc}^2 \,k_n \sdot k_c}
+(\e_n \sdot k_l) \sum_{c\neq l}^{n-1}
				\frac{\e_n\sdot k_l\,k_n\sdot k_c - 2 \,\e_n \sdot k_c\,k_n\sdot k_l}{\s_{lc}^2}  \\
& \qquad\qquad +k_l \sdot k_n
\left(
	\sum_{c\neq l}^{n-1} \frac{\e_n \sdot k_c}{\s_{lc}}
	-\frac{\e_n \sdot k_l}{k_n \sdot k_l}
	\sum_{c \neq l}^{n-1} \frac{k_n\sdot k_c}{\s_{lc}}
\right)^2\Bigg\}\d_{ml}\,.
\ea\ee
The first line in \eqref{Ieval} is the contribution of a double pole at $\s_n = \s_l$ and a double pole at $\s_n = \s_m$, whereas the second  line in \eqref{Ieval}  comes from the contribution of a single pole of the integrand at $\s_n = \s_c$, for all $c\neq l,m,n$. The first term in the third line comes from a single pole at $\s_n = \s_c$ for all $c\neq l,n $ and the remaining of \eqref{Ieval} comes from a third order pole at $\s_n = \s_l$.

\subsection{Evaluation of $m_2$}
For the evaluation of $m_2$ we need to expand \eqref{E_n} to order $\l$.  The derivative of the 
determinant of a $n\times n$ matrix with entries $T_{ab}$ can be obtained from the formula
\be\ba
\frac{d}{d\l}\det(T)&=\sum_{a=1}^n\sum_{b=1}^{n} (-1)^{a+b} \frac{dT_{ab}}{d\l} M^a_b,
\ea\ee
where $M^a_b$ denotes the determinant of the matrix obtained by removing the $a^{\rm th}$ row and the $b^{\rm th}$ column of $T$. Applying it onto $E_n$ in equation \eqref{E_n} yields
\be\ba \label{dE_nv1}
\frac{d E_n}{d\l} &=\sum_{a=1}^n\sum_{b=1}^n\left((-1)^{a+b}\frac{d A_{ab}}{d\l} \tilde \psi^a_b +2(-1)^{a+b+n}\frac{dC_{ab}}{d\l}\tilde \psi^{n+a}_b+(-1)^{a+b}\frac{d B_{ab}}{d\l}\tilde \psi^{n+a}_{n+b}\right).
\ea\ee
Here we have used the short notation
\be
\tilde{\psi}^a_b \equiv \frac{4 \,\det(\Psi^{12a}_{12b})}{\s_{12}^2} 
\d_{a\neq\{1,2\}}  \d_{b\neq\{1,2\}}.
\ee
For convenience and without loss of generality we have chosen to remove the first two rows and the first two columns in \eqref{E_n}. In \eqref{dE_nv1} we have also used the identity $\tilde \psi^a_b = -\tilde \psi^b_a$. The derivatives of the different matrix elements are
\be\ba
\frac{d A_{ab}}{d\l}&= \frac{1}{\s_{ab}}\left(\d_{an}k_n \sdot k_b +\d_{bn}k_a \sdot k_n\right)\d_{a\neq b} \, , \quad \frac{d B_{ab}}{d\l}=0\,,\\
&\qquad \frac{d C_{ab}}{d\l}=\frac{\e_a \sdot k_n}{\s_{ab}}\d_{bn}\d_{a\neq b}-\d_{ab}\frac{\e_a \sdot k_n}{\s_{an}}\d_{a\neq n}.
\ea\ee
Putting this into \eqref{dE_nv1} yields
\be\ba \label{dE_nv2}
\frac{d E_n}{d\l} &=2 \sum_{a=1}^{n-1}\frac{1}{\s_{na}}\left((-1)^{a+n}k_a \sdot k_n \, \tilde \psi^n_a + (-1)^a \e_a \sdot k_n \, \tilde \psi^n_{n+a} + (-1)^{n-1} \e_a \sdot k_n \, \tilde \psi^a_{n+a}\right).
\ea\ee
Note that all the dependence in $\l$ is now contained in the $\tilde \psi$ determinants only, which also need to be evaluated at $\l=0$ at the end. We further need to isolate any encounter of $\s_n$ in \eqref{dE_nv2}, since we eventually want to integrate over that variable.  We find
\be\ba 
\tilde{\psi}^n_a  &= C_{nn} \sum_{b=1}^{n-1} 
	\left( 
		(-1)^{n+b}\,\frac{\e_n \sdot k_b}{\s_{nb}} \psi^a_b
		-(-1)^b \, \frac{\e_n \sdot \e_b}{\s_{nb}} \psi^a_{n+b-1}
	\right),\\
\tilde{\psi}^n_{n+a}  &= -C_{nn} \sum_{b=1}^{n-1} 
	\left( 
		(-1)^{n+b}\,\frac{\e_n \sdot k_b}{\s_{nb}} \psi^{n+a-1}_b
		-(-1)^b \, \frac{\e_n \sdot \e_b}{\s_{nb}} \psi^{n+a-1}_{n+b-1}
	\right),\\
\tilde{\psi}_{n+a}^a &=- C_{nn}^2 \psi^a_{n+a-1},
\ea\ee
where we have dropped the tilde sign to denote the further removal of the rows and columns that contain the variable $\s_n$, that is $\psi^a_b$ denotes the determinant $E_{n-1}$ after the removal of the $a^{\rm th}$ row and the $b^{\rm th}$ column.  Then
\be\ba
E_n^{(1)} &= 2 C_{nn}\sum_{a=1}^{n-1}\sum_{b=1}^{n-1} \frac{(-1)^{a+b}}{\s_{na}\s_{nb}}
\Big( 
	k_n \sdot k_a\,\e_n \sdot k_b \,\psi^a_b
	-k_n \sdot \e_b\,\e_n \sdot \e_a\, \psi^{n+a-1}_{n+b-1} 
 \\
&\qquad\qquad 
	+(-1)^n\left(\e_n \sdot k_a\,k_n \sdot \e_b-k_n \sdot k_a\,\e_n \sdot \e_b\right)\psi^a_{n+b-1}
\Big)\\
&\quad +2C_{nn}^2\sum_{a=1}^{n-1} \frac{(-1)^n}{\s_{na}}\e_a \sdot k_n \psi^a_{n+a-1}.
\ea\ee
We recall that $m_2$ takes the form
\be\ba
m_2 &=\int d\s_n \d(f_n^{n-1})\sum_{l\neq i,j,k}^{n-1}\frac{k_n \sdot k_l}{\s_{ln}} \d'(f_l^{n-1})\prod_{m\neq i,j,k,l}^{n-1}\d(f_m^{n-1})E_n^{(1)} 
\ea\ee
thus, we will need the following integrals
\be\ba
I_3 \equiv - k_n \sdot k_l\int d\s_n  \d(f_n^{n-1}) \frac{C_{nn}}{\s_{nl}\s_{na}\s_{nb}}\label{I3},\\
\ea\ee
\be\ba
I_4 \equiv - k_n \sdot k_l \int d\s_n  \d(f_n^{n-1}) \frac{C_{nn}^2}{\s_{nl}\s_{na}}.\label{I4}
\ea\ee
The integral $I_4$ is directly obtained from \eqref{Ieval} since $I_4=-(k_n \sdot k_a)^{-1} I|_{m=a}$. For $I_3$ we find
\be\ba
\label{I3_eval}
I_3&=k_n \sdot k_l\left[\frac{\e_n \sdot k_a}{k_n \sdot k_a}\frac{1}{\s_{al}\s_{ab}}+(a\leftrightarrow l)+(a\leftrightarrow b)\right]\d_{l\neq a}\d_{l\neq b}\d_{a\neq b}\\
&\quad+k_n \sdot k_l \Bigg[\left(\frac{\e_n \sdot k_a}{k_n \sdot k_a}\frac{1}{\s_{al}}\sum_{c\neq a}^{n-1}\frac{1}{\s_{ac}}\left(\frac{\e_n \sdot k_c}{\e_n \sdot k_a}-\frac{k_n \sdot k_c}{k_n \sdot k_a}\right)+\frac{\e_n \sdot k_l}{k_n \sdot k_l}\frac{1}{\s_{al}^2}-\frac{\e_n \sdot k_a}{k_n \sdot k_a}\frac{1}{\s_{al}^2}\right)\d_{l\neq a}\d_{ab}\\
&\hspace{80pt}+\left(l\leftrightarrow a\right)+(\{a,l,b\} \to \{l,b,a\})\Bigg]\\
&\quad+\e_n \sdot k_l \left[-\frac{1}{\e_n \sdot k_l}\sum_{c\neq l}^{n-1}\frac{\e_n \sdot k_c}{\s_{lc}^2}-\frac{1}{k_n \sdot k_l \e_n \sdot k_l}\sum_{c\neq l}^{n-1}\sum_{d\neq l}^{n-1}\frac{k_n \sdot k_c \e_n \sdot k_d}{\s_{lc}\s_{ld}}\right.\\
&\hspace{80pt} \left.+\frac{1}{k_n \sdot k_l}\sum_{c\neq l}^{n-1}\frac{k_n \sdot k_c}{\s_{lc}^2}+\frac{1}{(k_n \sdot k_l)^2}\left(\sum_{c\neq l}^{n-1}\frac{k_n \sdot k_c}{\s_{lc}}\right)^2\right]\d_{ab}\d_{bl}.
\ea\ee
As a check, note that from this expression the quantity $I_3/(k_n \sdot k_l)$ is symmetric under the exchange of any two pairs of $(l,a,b)$ which is evident from the original definition in \eqref{I3}. 

We now write $m_2$ making explicit the linear combination of the different types of minors we have, i.e.,
\be\ba\label{m2lin1}
m_2 = 2\sum_{l\neq i,j,k}^{n-1}\d'(f_l^{n-1})\prod_{m\neq i,j,k,l}^{n-1}\d(f_m^{n-1})D_l,
\ea\ee 
where
\be\ba
D_l\equiv &\sum_{a\neq l}^{n-1}\sum_{b\neq l,a}^{n-1} \left(c_1\psi^a_b +c_2 \psi^{n+a-1}_{n+b-1}+c_3\psi^{a}_{n+b-1}\right)I_3{}_{\{l\neq a,\, l \neq b, \,a\neq b\}}\\
&+\sum_{a\neq l}^{n-1} c_4\psi^a_{n+a-1}+ \sum_{a\neq l}^{n-1}\left(c_5\psi^a_l + c_6\psi^{n+a-1}_{n+l-1}+ c_7\psi^a_{n+l-1}+c_8\psi^l_{n+a-1}\right)I_3{}_{\{l=b,\, l\neq a \}}\\
&+c_9\psi^l_{l+n-1}\label{detlincomb}.
\ea\ee
The coefficients $c_i$ are
\be\ba
\label{m2_coeff}
c_1 &= (-1)^{a+b}k_n \sdot k_a \, \e_n \sdot k_b\,;\hspace{5pt} 
c_2=-  (-1)^{a+b} \e_b \sdot k_n \,\e_n \sdot \e_a \,;\hspace{5pt}
c_3= (-1)^n \left(\e_n \sdot k_a\, \e_b \sdot k_n -k_n \sdot k_a \, \e_n \sdot \e_b\right);\\
c_4&= (-1)^n\left(\e_n \sdot k_a \,k_n \sdot \e_a -k_n \sdot k_a \,\e_n \sdot \e_a\right)I_3{}_{\{a=b, \,l\neq a \}}+(-1)^n \e_a \sdot k_n \,I_4{}_{\{l\neq a \}};\\
c_5&= (-1)^{a+l}\left(k_n \sdot k_a \,\e_n \sdot k_l - k_n \sdot k_l \,\e_n \sdot k_a \right) 
\,;\hspace{5pt}
c_6= (-1)^{a+l}\left(\e_a \sdot k_n \,\e_n \sdot \e_l - \e_l \sdot k_n \,\e_n \sdot \e_a \right);\\
c_7&= (-1)^{a+l+n}\left(\e_n \sdot k_a \,\e_l \sdot k_n - k_n \sdot k_a \,\e_n \sdot \e_l \right)
\,;\hspace{5pt}
c_8= (-1)^{a+l+n}\left(\e_n \sdot k_l \,\e_a \sdot k_n - k_n \sdot k_l \,\e_n \sdot \e_a \right);\\
c_9&=  (-1)^n\left(\e_n \sdot k_l \, k_n \sdot \e_l -k_n \sdot k_l \,\e_n \sdot \e_l\right)I_3{}_{\{l=a=b\}}+(-1)^n \e_l \sdot k_n \,I_4{}_{\{l=a \}}.
\ea\ee
In the above we have used the identity $\psi^a_b = - \psi^b_a$.

\subsection{Evaluation of $m_3$} 
We define $\tilde \psi^{ab}_{cd}$ and  $\psi^{ab}_{cd}$ to be respectively the determinants $E_{n}$ and $E_{n-1}$ after the removal of the rows $a,b$ and the columns $c,d$.

For the evaluation of $m_3$ we need to take the second derivative of \eqref{E_n} with respect to $\l$.  From \eqref{dE_nv2} we have
\be\ba \label{dE_nv3}
\frac{d^2 E_n}{d\l^2} &=2 \sum_{a=1}^{n-1}\frac{1}{\s_{na}}\left(
(-1)^{a+n}k_a \sdot k_n \, \frac{d \tilde \psi^n_a}{d\l} 
+ (-1)^a \e_a \sdot k_n \, \frac{d \tilde \psi^n_{n+a}}{d\l} 
+ (-1)^{n-1} \e_a \sdot k_n \, \frac{d \tilde \psi_{n+a}^a}{d \l}\right).
\ea\ee
With the definition $\theta_{ij}$ to be 0 when $i>j$ and $-1$ when $i<j$ we find
\be\ba\label{dpsidl}
\frac{d \tilde \psi^n_a}{d\l} &=\sum_{b=1}^{n-1} \frac{1}{\s_{bn}} \left(
	 (-1)^{n+b-1} k_b \sdot k_n \tilde \psi^{bn}_{an}
	+ (-1)^{n-1} \e_b \sdot k_n \tilde \psi ^{bn}_{a,n+b}
	+(-1)^b \e_b \sdot k_n \tilde \psi^{n,n+b}_{an} \right) \\
&\hspace{2cm}+\sum_{b\neq a}^{n-1} (-1)^{n+ \theta_{ab}} \frac{\e_b \sdot k_n}{\s_{bn}} \tilde \psi^{n,n+b}_{ab}\,,\\
 \frac{d \tilde \psi^n_{n+a}}{d\l} &=\sum_{b=1}^{n-1} \frac{1}{\s_{bn}} \left(
	(-1)^{n+b} k_b \sdot k_n \tilde \psi^{bn}_{n,n+a}
	+(-1)^{b-1} \e_b \sdot k_n \tilde \psi^{n,n+b}_{n,n+a} 
	+ (-1)^n \e_b \sdot k_n \tilde \psi^{n,n+b}_{b,n+a}\right) \\
&\hspace{2cm}+\sum_{b\neq a}^{n-1} (-1)^{n+ \theta_{ab}} \frac{\e_b \sdot k_n}{\s_{bn}} \tilde \psi^{bn}_{n+a,n+b}\,,\\
\frac{d \tilde \psi_{n+a}^a}{d \l} &=\sum_{b=1}^{n-1} \frac{1}{\s_{bn}} \left(
	(-1)^{n+b} k_b \sdot k_n \tilde \psi^{an}_{b,n+a}
	+(-1)^{n} \e_b \sdot k_n \tilde \psi^{a,n+b}_{b,n+a} 
	+ (-1)^{b-1} \e_b \sdot k_n \tilde \psi^{a,n+b}_{n,n+a}\right)\\
 &\hspace{-0.5cm}+\sum_{b\neq a}^{n-1} \frac{(-1)^{\theta_{ab}}}{\s_{bn}} \left(
	(-1)^{n+b} k_b \sdot k_n \tilde \psi^{ab}_{n,n+a}
	+(-1)^{n+\theta_{ab}} \e_b \sdot k_n \tilde \psi^{ab}_{n+a,n+b} 
	+ (-1)^{b} \e_b \sdot k_n \tilde \psi^{an}_{n+a,n+b}\right).
\ea\ee
We can further expand the $n$ and $2n$ rows and columns of the minors appearing in \eqref{dpsidl}.  With the help of the identity $\psi^{ab}_{cd} =\psi^{cd}_{ab}$ we arrive at the following result
\be
E_n^{(2)} = C_{nn}^2 A_1 + C_{nn} A_2 + A_3,
\ee
where
\be
A_1 = \sum_{a=1}^{n-1} \sum_{b=1}^{n-1} \frac{\e_a \sdot k_n \e_b \sdot k_n}{\s_{na}\s_{nb}} 
\psi^{a,n+b-1}_{b,n+a-1}
+\sum_{a=1}^{n-1} \sum_{b \neq a }^{n-1} \frac{\e_a \sdot k_n \e_b \sdot k_n}{\s_{na}\s_{nb}} 
\psi^{ab}_{n+a-1,n+b-1},
\ee
\be\ba
A_2 =&2 \sum_{a=1}^{n-1} \sum_{b=1}^{n-1} \sum_{c=1}^{n-1} \frac{(-1)^{b+c}}{\s_{na}\s_{nb}\s_{nc}}
 \e_a \sdot k_n (\e_b \sdot k_n \e_n \sdot k_c - k_c \sdot k_n \e_b \e_n) \psi^{a,n+b-1}_{c,n+a-1}\\
+&2 \sum_{a=1}^{n-1} \sum_{c=1}^{n-1} \sum_{b\neq a}^{n-1} \frac{(-1)^{b+c+n+\theta_{ab}} }{\s_{na}\s_{nb}\s_{nc}}
\e_a \sdot k_n \\
&\hspace{1cm}\left[ (k_b \sdot k_n \e_n \sdot k_c - k_c k_n \e_n \sdot k_b) \psi^{ab}_{c,n+a-1}
+(\e_b \sdot k_n \e_c \sdot \e_n -\e_c \sdot k_n \e_b \sdot \e_n  ) \psi^{a,n+c-1}_{n+a-1,n+b-1}
\right] \\
+&2 \sum_{a=1}^{n-1} \sum_{b \neq a}^{n-1} \sum_{c\neq a}^{n-1} 
\frac{(-1)^{b+c+\theta_{ab}+\theta_{ac}} }{\s_{na}\s_{nb}\s_{nc}}
\e_a \sdot k_n 
(\e_c \sdot k_n \e_n \sdot k_b - k_b \sdot k_n \e_c \sdot \e_n) \psi^{ab}_{n+a-1,n+c-1},
\ea\ee
\be\ba
A_3 =&\sum_{a=1}^{n-1} \sum_{b=1}^{n-1} \sum_{c = 1}^{n-1}\sum_{d = 1}^{n-1}
\frac{(-1)^{a+b+c+d}}{\s_{na}\s_{nb}\s_{nc}\s_{nd}}
\left(
	\e_n \sdot k_a \e_n \sdot k_b \e_c \sdot k_n \e_d \sdot  k_n
	+k_a \sdot k_n k_b \sdot k_n \e_c \sdot \e_n \e_d \sdot \e_n
\right. \\
&\hspace{7cm}
\left.
	-2 k_a \sdot k_n \e_n \sdot k_b \e_c \sdot k_n \e_d \sdot \e_n
\right)
\psi^{b,n+c-1}_{a,n+d-1}\\
+2&\sum_{a=1}^{n-1} \sum_{b=1}^{n-1} \sum_{c = 1}^{n-1}\sum_{d \neq a}^{n-1}
\frac{(-1)^{a+b+c+d+n+\theta_{ad}}}{\s_{na}\s_{nb}\s_{nc}\s_{nd}}
(
	k_a \sdot k_n \e_n \sdot k_d (k_b \sdot k_n \e_c \sdot \e_n- \e_n \sdot k_b \e_c \sdot k_n)
	\psi_{b,n+c-1}^{ad}
\\
&  \hspace{4.5cm}
	+\e_a \sdot k_n \e_d \sdot \e_n (k_b \sdot k_n \e_c \e_n -\e_n \sdot k_b \e_c \sdot k_n)
	\psi^{b,n+c-1}_{n+a-1,n+d-1}
)\\
+&\sum_{a=1}^{n-1} \sum_{c=1}^{n-1} \sum_{b \neq c}^{n-1}\sum_{d \neq a}^{n-1}
\frac{(-1)^{a+b+c+d+\theta_{ad}+\theta_{cb}}}{\s_{na}\s_{nb}\s_{nc}\s_{nd}}
(
	2 k_a \sdot k_n \e_b \sdot \e_n \e_c \sdot k_n \e_n \sdot k_d
	\psi^{ad}_{n+b-1,n+c-1}
\\
& \hspace{3cm}
        +k_a \sdot k_n \e_n \sdot k_b k_c \sdot k_n  \e_n \sdot k_d \psi^{bc}_{ad}
        + \e_a \sdot k_n \e_b \sdot \e_n \e_c \sdot k_n  \e_d \sdot \e_n\psi^{n+b-1,n+c-1}_{n+a-1,n+d-1}
).
\ea\ee

In order to finish the calculation of $m_3$ the only new integral we need to evaluate is
\be
I_5 \equiv \int d\s_n \d(f_n^{n-1})\frac{1}{\s_{na}\s_{nb}\s_{nc}\s_{nd}}
\ee
for which we obtain
\be\ba
I_5 &=\frac{1}{(k_n \sdot k_a)^2}\left[\sum_{l\neq a}^{n-1}\frac{k_n \sdot k_l}{\s_{al}^2}
+\frac{1}{k_n \sdot k_a}\left(\sum_{l\neq a}^{n-1} \frac{k_n \sdot k_l}{\s_{al}}{}\right)^2\right]\d_{ab}\d_{bc}\d_{cd}\\
&\quad + \Bigg\{ \frac{1}{k_n \sdot k_b} \frac{1}{\s_{ab}}\left[\frac{1}{k_n \sdot k_b}\sum_{l\neq b}^{n-1}\frac{k_n \sdot k_l}{\s_{bl}}-\frac{1}{\s_{ab}}\right]\d_{a\neq b}\d_{bc}\d_{cd} + {\rm cyclic}\, \{a,b,c,d\} \Bigg\}\\
&\quad  + \Bigg\{ \left(\frac{1}{k_n \sdot k_a}+\frac{1}{k_n \sdot k_c}\right)\frac{1}{\s_{ac}^2}\d_{ab}\d_{cd}\d_{a \neq c}+ {\rm cyclic}\{b,c,d\}\Bigg\}\\
&\quad + \Bigg\{\frac{1}{\s_{ac}\s_{ad}}\frac{1}{k_n \sdot k_a}\d_{ab}\d_{c\neq d}\d_{a\neq c,d} + {\rm cyclic}\{b,c,d\}\Bigg\}\\
&\quad + \Bigg\{  \frac{1}{\s_{ba}\s_{bd}}\frac{1}{k_n \sdot k_b} \d_{bc}\d_{a\neq d}\d_{b\neq a,d}+ {\rm cyclic}\{b,c,d\} \Bigg\}.
\ea\ee

\section{Action of $S^{(2)}$ on the amplitude}
From \eqref{S2}, the complete expression for $S^{(2)}$ including the spin contribution can be written as
\be\label{s^2}
S^{(2)}=S^{(2)}_{\rm orb} + S^{(2)}_{\rm so} +S^{(2)}_{\rm spin},
\ee
where the orbital, spin-orbit and spin parts are respectively given by
\be\ba\label{ss^2}
S^{(2)}_{\mathrm{orb}} =& \frac 1 2  \sum_{a=1}^{n-1}K_{a \mu\nu}^{\rm orb}\,
\frac{\partial^2}{\partial k_{a\m} \partial k_{a\n}},\quad 
S^{(2)}_{\rm so}=\sum_{a=1}^{n-1}K_{a \mu\nu}^{\rm so}\,\frac{\pd^2}{\pd k_{a\m}\pd \e_{a\n}},\quad
S^{(2)}_{\rm spin}= \frac 1 2  \sum_{a=1}^{n-1} K_{a \mu\nu}^{\rm spin}\,\frac{\pd^2}{\pd \e_{a\m}\pd \e_{a\n}},
\ea\ee
with
\be\ba\label{Ks}
K_{a \mu\nu}^{\rm orb}\equiv & 
 k_n\sdot k_a \,\e_{n\m} \e_{n\n} -\e_n \sdot k_a\, \e_{n(\m} 
k_{n\n)} + \frac{(\e_n \sdot k_a)^2}{k_n \sdot k_a} k_{n\m} k_{n\n} \,,\\
K_{a \mu\nu}^{\rm so}\equiv &
\e_a \sdot k_n\, \e_{n\m} \e_{n\n} -\e_n \sdot \e_a \,\e_{n\m} 
k_{n\n}-\frac{\e_n \sdot k_a\,\e_a \sdot k_n}{k_n \sdot k_a}\e_{n\n} k_{n\m} 
+ \frac{\e_n \sdot k_a\,\e_n \sdot \e_a}{k_n\sdot k_a} k_{n\m}k_{n\n}\,,\\
K_{a \mu\nu}^{\rm spin}\equiv &
\frac{(\e_{a} \sdot k_n)^2}{k_n\sdot k_a}\e_{n\m} \e_{n\n} -\frac{\e_a \sdot k_n\,\e_n \sdot \e_a}{k_n \sdot k_a}\e_{n(\m} k_{n\n)}+\frac{(\e_n \sdot \e_a)^2}{k_n \sdot k_a}k_{n\mu}k_{n\n}\,.
\ea\ee

Then the action of $S^{(2)}$ on the amplitude is
\be\ba
\label{fullS2}
S^{(2)} M_{n-1} &= S^{(2)} \int [d\s]_{n-4} \prod_{l\neq i,j,k}^{n-1} \hspace{-5pt}\delta(f_l^{n-1}) E_{n-1} \\
&= \int [d\s]_{n-4} \left(s_1 +s_2+s_3+s_4 \right),
\ea\ee
where we have separated the calculation into the following four parts
\be\ba\label{s's}
s_1 &= E_{n-1} \,S^{(2)}_{\mathrm{orb}} \prod_{l\neq i,j,k}^{n-1} \hspace{-5pt}\delta(f_l^{n-1}), \qquad 
s_2 = \sum_{a=1}^{n-1} K_{a \mu\nu}^{\rm orb}
  \frac{\partial E_{n-1}}{\partial k_{a\m}}
  \frac{\partial}{\partial k_{a\n}} \prod_{l\neq i,j,k}^{n-1} \hspace{-5pt} \delta(f_l^{n-1}), \\
s_3 &= \sum_{a=1}^{n-1} K_{a \mu\nu}^{\rm so}
\frac{\partial E_{n-1}}{\partial \e_{a\n}}
  \frac{\partial}{\partial k_{a\m}} \prod_{l\neq i,j,k}^{n-1} \hspace{-5pt} \delta(f_l^{n-1}), \qquad
s_4 = \prod_{l\neq i,j,k}^{n-1} \hspace{-5pt}\delta(f_l^{n-1}) \, S^{(2)} E_{n-1}.
\ea\ee

In the subsequent computations we will make use of the identities
\be\ba\label{identities}
\frac{\pd f_l^{n}}{\pd k_{a\m}}=\frac{k_l^{\m}}{\sigma_{la}}\delta_{l\neq a}+ 
\delta_{la}\sum_{d\neq l}^n\frac{k_d^{\m}}{\sigma_{ld}}\, ,\qquad 
\frac{\pd^2 f_l^{n}}{\pd k_{a\m}\pd k_{a\n}}=0,
\ea\ee
and also 
\be \ba\label{dEdl}
\frac{\partial E_{n-1}}{\partial k_{a\mu}} &= 2\sum_{b\neq a}^{n-1} \frac{1}{\s_{ab}}
	\left(
		(-1)^{a+b} {k_b}^{\m} \psi^a_b +(-1)^{a+b+n+1} {\e_b}^{\mu} \psi^a_{n+b-1} +(-1)^{n} {\e_b}^{\mu} 	\psi^b_{n+b-1} 
	\right) ,\\
\frac{\partial E_{n-1}}{\partial \e_{a\mu}} &=	2\sum_{b\neq a}^{n-1} \frac{1}{\s_{ab}}
	\left(
		(-1)^{a+b+n} {k_b}^{\mu} \psi^{b}_{n+a-1} + (-1)^{n+1} {k_b}^{\mu} \psi^{a}_{n+a-1}
			+(-1)^{a+b}{\e_b}^{\mu} \psi^{n+a-1}_{n+b-1}
	\right).
\ea \ee
In the following we omit the upper index of the scattering equations $f_l^{n-1}$ and we simply write them as $f_l$.

\subsection{Evaluation of $s_1$} 
We find
\be\ba
s_1 &= \frac 1 2 \, E_{n-1} \hspace{-5pt} \sum_{l\neq i,j,k}^{n-1}\hspace{0pt} \d'(f_l)\hspace{-5pt}
				\sum_{m\neq i,j,k,l}^{n-1}\hspace{0pt} \d'(f_m)\hspace{-5pt}
				\prod_{b\neq i,j,k,l,m}^{n-1}\hspace{-5pt} \d(f_b)
				\sum_{a=1}^{n-1} K_{a\m\n}^{\rm orb} \frac{\partial f_m}{\partial k_{a\m}}
					\frac{\partial f_l}{\partial k_{a\n}}\\
	&\hspace{100pt} + \frac 1 2 \, E_{n-1} \hspace{-5pt} \sum_{l \neq i,j,k}^{n-1} \hspace{-5pt}\d''(f_l)
                \prod_{b\neq i,j,k,l}^{n-1} \hspace{-5pt} \d(f_b)
                \sum_{a=1}^{n-1} K_{a\m\n}^{\rm orb} \frac{\partial f_l}{\partial k_{a\m}}
					\frac{\partial f_l}{\partial k_{a\n}}.
\ea\ee
After some straightforward algebra and using \eqref{Ks} and \eqref{identities} we obtain  
\be\ba 
&\sum_{a=1}^{n-1} K_{a\m\n}^{\rm orb} \frac{\partial f_m}{\partial k_{a\m}}
					\frac{\partial f_l}{\partial k_{a\n}} =  
					\d_{ml} \, I_2 +  \d_{m \neq l}\, I_1.
\ea\ee
thus, comparing with \eqref{m1}, we obtain the desired result $s_1=m_1$.

\subsection{Evaluation of $s_2$ and $s_3$} 
The combination $s_2+s_3$ has the same delta function support as $m_2$, thus, we will compare these two expressions. For $s_2$ we obtain
\be\ba
s_2 &=  \sum_{l\neq i,j,k}^{n-1}\d'(f_l)\prod_{m\neq i,k,j,l}^{n-1}\d(f_m)\sum_{a=1}^{n-1} K_{a\m\n}^{\rm orb} 
\frac{\pd E_{n-1}}{\pd k_{a\m}}\frac{\pd f_l}{\pd k_{a\n}}
\ea\ee
and for $s_3$ we get
\be\ba
s_3 &= \sum_{l\neq i,j,k}^{n-1}\d'(f_l)\prod_{m\neq i,k,j,l}^{n-1}\d(f_m)\sum_{a=1}^{n-1}K_{a\m\n}^{\rm so}\frac{\pd E_{n-1}}{\pd \e_{a\n}}\frac{\pd f_l}{\pd k_{a\m}}.
\ea\ee
After some tedious but straightforward algebra and using \eqref{Ks},  \eqref{identities} and \eqref{dEdl} we can expand $s_2+s_3$ in the same form of $m_2$ as shown in \eqref{m2lin1} and \eqref{detlincomb}. We have explicitly computed each of the coefficients of the corresponding expansion for $s_2 + s_3$ and see that they all precisely match those of \eqref{m2_coeff}, thus, arriving at $s_2+s_3=m_2$ as expected.

\subsection{Evaluation of $s_4$} 
Having matched all the previous terms on both sides, our last task is to show that $s_4=m_3$.

From \eqref{s's}, \eqref{s^2} and \eqref{ss^2} we have
\be
s_4 = \prod_{l\neq i,j,k}^{n-1} \hspace{0pt}\delta(f_l^{n-1}) \,
\sum_{a=1}^{n-1} \left[
\frac 1 2 K_{a\m\n}^{\rm orb} \frac{\partial^2 E_{n-1}}{\partial k_{a\n} \partial k_{a\m}} 
+ K_{a\m\n}^{\rm so} \frac{\partial^2 E_{n-1}}{\partial \e_{a\n} \partial k_{a\m}} 
+\frac 1 2 K_{a\m\n}^{\rm spin} \frac{\partial^2 E_{n-1}}{\partial \e_{a\n} \partial \e_{a\m}} 
\right].
\ee
Appropriately differentiating \eqref{dEdl} we find
\be\ba
\frac{\partial^2 E_{n-1}}{\partial k_{a\n} \partial k_{a\m}}  =&
2\sum_{b\neq a}^{n-1} \sum_{c\neq a}^{n-1} \frac 1 {\s_{ab}\s_{ca}}
\left(
(-1)^{b+c+\theta_{ba}+\theta_{ac}} {k_b}^{\m}{k_c}^{\n}\psi^{ac}_{ab}
\right. \\
& \left. \hspace{-1.5cm}
+(-1)^{a+b+n+\theta_{ac}} ({k_b}^{\m}{\e_c}^{\n} + {\e_c}^{\m}{k_b}^{\n}) \psi^{ac}_{b,n+c-1}
+(-1)^{b+c+n+\theta_{ba}} ({k_b}^{\m}{\e_c}^{\n} + {\e_c}^{\m}{k_b}^{\n}) \psi^{ab}_{a,n+c-1} \right. \\
& \left. \hspace{-1.5cm}
+(-1)^{a+b} ({\e_b}^{\m}{\e_c}^{\n} + {\e_c}^{\m}{\e_b}^{\n} ) \psi^{a,n+c-1}_{c,n+b-1}
- {\e_b}^{\m}{\e_c}^{\n} ( \psi^{b,n+c-1}_{c,n+b-1} +(-1)^{b+c} \psi^{a,n+c-1}_{a,n+b-1} )
\right)\\
+&2\sum_{b\neq a}^{n-1} \sum_{c\neq a,b}^{n-1} \frac 1 {\s_{ab}\s_{ca}}
\left(
(-1)^{a+b+n+\theta_{cb}} ({k_b}^{\m}{\e_c}^{\n} +{\e_c}^{\m}{k_b}^{\n}) \psi^{bc}_{a,n+c-1}
\right. \\
& \left. \hspace{-1.5cm}
+(-1)^{a+b+\theta_{bc}+\theta_{ac}} ( {\e_b}^{\m}{\e_c}^{\n} + {\e_c}^{\m}{\e_b}^{\n} ) \psi^{ac}_{n+b-1,n+c-1}
- {\e_b}^{\m}{\e_c}^{\n} \psi^{bc}_{n+b-1,n+c-1}
\right),
\ea\ee
\be\ba
\frac{\partial^2 E_{n-1}}{\partial \e_{a\n} \partial k_{a\m}}  =&
2\sum_{b\neq a}^{n-1} \sum_{c\neq a}^{n-1} \frac 1 {\s_{ab}\s_{ac}}
\left[
(-1)^{b+n} {k_b}^{\m} {k_c}^{\n}((-1)^{c+\theta_{ca}} \psi^{ac}_{b,n+a-1} +(-1)^{a+\theta_{ab}} \psi^{ab}_{a,n+a-1})
\right. \\
&\left. \hspace{-1.5cm}
+(-1)^{b+c}( {k_b}^{\m} {\e_c}^{\n} - {\e_c}^{\m} {k_b}^{\n})\psi^{a,n+a-1}_{b,n+c-1}
-(-1)^{b+c}  {k_b}^{\m} {\e_c}^{\n} \psi^{a,n+c-1}_{b,n+a-1}
\right. \\
&\left. \hspace{-1.5cm}
+{\e_b}^{\m} {k_c}^{\n}
((-1)^{b+c+\theta_{ab}+\theta_{ac}} \psi^{ac}_{n+a-1,n+b-1} - \psi^{ab}_{n+a-1,n+b-1})
\right. \\
&\left. \hspace{-1.5cm}
+{\e_b}^{\m} {k_c}^{\n}
((-1)^{a+b} \psi^{a,n+a-1}_{a,n+b-1}
+(-1)^{a+c} \psi^{b,n+a-1}_{c,n+b-1}
-\psi^{b,n+a-1}_{a,n+b-1})
\right. \\
&\left. \hspace{-1.5cm}
+{\e_b}^{\m} {\e_c}^{\n}
(-1)^{c+n+\theta_{ab}}
((-1)^b \psi^{a,n+c-1}_{n+a-1,n+b-1}
-(-1)^a \psi^{b,n+c-1}_{n+a-1,n+b-1})
\right] \\
+&2\sum_{b\neq a}^{n-1} \sum_{c\neq a,b}^{n-1} \frac 1 {\s_{ab}\s_{ac}}
\left[
(-1)^{b+c+n+\theta_{bc}} {k_b}^{\m} {k_c}^{\n} \psi_{a,n+a-1}^{bc} \right.\\
&\left. \hspace{-1.5cm}
+(-1)^{a+c+n+\theta_{cb} } {\e_b}^{\m} {\e_c}^{\n} \psi^{b,n+a-1}_{n+b-1,n+c-1}
+(-1)^{b+c+n+\theta_{bc}} {\e_b}^{\m} {\e_c}^{\n} \psi^{a,n+a-1}_{n+b-1,n+c-1} 
\right. \\
&\left. \hspace{-1.5cm}
+(-1)^{a+c+\theta_{ba}+\theta_{bc}}  {\e_b}^{\m} {k_c}^{\n} \psi^{bc}_{n+a-1,n+b-1}
\right],
\ea\ee
\be\ba
\frac{\partial^2 E_{n-1}}{\partial \e_{a\n} \partial \e_{a\m}}  =&
2\sum_{b\neq a}^{n-1} \sum_{c\neq a}^{n-1} \frac 1 {\s_{ab}\s_{ac}}
\left[
 {k_b}^{\m} {k_c}^{\n} \left((-1)^{b+c} \psi^{b,n+a-1}_{c,n+a-1} 
+ \psi^{a,n+a-1}_{a,n+a-1}\right)
\right. \\
&\left. \hspace{-1.5cm}
+ (-1)^{b+c+\theta_{ba}+\theta_{ca}} {\e_b}^{\m} {\e_c}^{\n} \psi^{n+a-1,n+b-1}_{n+a-1,n+c-1}
-(-1)^{a+b}  ({k_b}^{\m} {k_c}^{\n}+{k_c}^{\m} {k_b}^{\n}) \psi^{a,n+a-1}_{b,n+a-1} \right. \\
&\left. \hspace{-1.5cm}
+(-1)^{c+n+\theta_{ca}} ({k_b}^{\m} {\e_c}^{\n}+{\e_c}^{\m} {k_b}^{\n})
\left((-1)^b  \psi^{b,n+a-1}_{n+a-1,n+c-1} - (-1)^a \psi^{a,n+a-1}_{n+a-1,n+c-1}\right)
\right].
\ea\ee
We now have all the ingredients to perform the comparison of $s_4$ with $m_3$.  The algebra is tedious but straightforward since it only involves changes of the summation order and renaming dummy indices.  We have performed the analysis and found agreement of the two expressions which completes the proof of the soft-graviton theorem.

\vspace{3mm}

\noindent
{\bf Acknowledgments}

\vspace{3mm}

\noindent
We thank Wei He for useful discussions.  We also thank Anastasia Volovich and Michael Zlotnikov for correspondence. In the final stages of the writing of this paper we learned of the work of Michael Zlotnikov \cite{MZ} who was also embarked on our same computations. C.K. would like to thank the organizers of the Simons workshop during which this project was completed.  The work of C.K. is supported by the S\~ao Paulo Research Foundation (FAPESP) under grants 2011/11973-4 and 2012/00756-5. The work of F.R is supported by FAPESP grant 2012/05451-8.

\bibliographystyle{utphys}		
\bibliography{mybib}
\end{document}